# Electric Charge Transport and Dielectric Properties of the Barium Titanate Ceramics Obtained by Spark-Plasma Sintering with Different Carbon Content

Oleksandr S. Pylypchuk[1*], Victor V. Vainberg[1†], Denis O. Stetsenko[1], Oleksii V. Bereznikov[1], Taisiia O. Kuzmenko[1,2], Serhii E. Ivanchenko[3], Bohdan Pokhylko[3], Vladyslav Kushnir[3], Lesya Demchenko[4,5], Volodimir N. Poroshin[1], Victor I. Styopkin [1‡], and Anna N. Morozovska[1§]

[1] Institute of Physics of the National Academy of Sciences of Ukraine,
46 Nauky Avenue, 03028 Kyiv, Ukraine

[2] Educational Scientific Institute of High Technologies, Taras Shevchenko National University of Kyiv, 4-g Hlushkova Avenue, 03022 Kyiv, Ukraine

[3] Frantsevich Institute for Problems in Materials Science, National Academy of Sciences of Ukraine Omeliana Pritsaka str., 3, Kyiv, 03142, Ukraine

[4] Department of Chemistry, Stockholm University, Stockholm 10691, Sweden

[5] Department of Physical Materials Science and Heat Treatment, National Technical University of Ukraine "Igor Sikorsky Kyiv Polytechnic Institute", Kyiv 03056, Ukraine

**Abstract**

Barium titanate (BaTiO$_3$) ceramics with a different content of carbon were synthesized by spark-plasma sintering (SPS) at the temperature of 1100°C in vacuum under pressure. The concentration and distribution of carbon impurity inside the samples was estimated by scanning electron microscopy. The resistivity vs temperature and electric field dependences of the SPS ceramics with different carbon concentration have been studied. It is shown that their conduction is determined by the variable range hopping mechanism and obeys the Mott law. The density of localized states and localization radius of the electron wave function are determined. The difference in low-temperature resistivity of the SPS ceramics is caused by carbon concentration and connected with it variation of the dielectric permittivity. The relative dielectric permittivity of the SPS ceramics is colossal and reaches the values of $10^5 - 10^6$ order. The larger carbon concentration is, the smaller the permittivity and resistivity are within the Mott hopping conduction temperature range. In the range from 250 K to 408 K one observes that the dielectric permittivity strongly increases forming a maximum in all samples, which may be related to the phase transition. Along with this, resistivity manifests a simultaneous sharp decrease. The decrease of resistivity along with the characteristic dependence of resistivity vs dielectric permittivity in the Mott conduction temperature range, evidences the validity of Heywang model for the description of SPS ceramics conduction mechanisms. The resistivity strongly decreases with increasing frequency in the AC regime, which agrees both with models of hopping conduction and effects based on the Maxwell-Wagner model. The studied SPS $BaTiO_3$ ceramics are attractive for applications in energy storage and sensorics.

---

[*] Corresponding author: alexander.pylypchuk@gmail.com
[†] Corresponding author: viktor.vainberg@gmail.com
[‡] Corresponding author: vstyopkin@gmail.com
[§] Corresponding author: anna.n.morozovska@gmail.com

## 1. INTRODUCTION

The importance of ferroelectric ceramics for usage in capacitors and supercapacitors is conditioned by their colossal relative dielectric permittivity, which can reach $10^5 - 10^6$ and more [1, 2, 3]. Using nano-grained ferroelectric ceramics may be advantageous for electrostatic supercapacitors [4] and energy storage devices due to the negative capacitance state [5] emerging in nanoscale ferroelectric films [6, 7] and nanoparticles [8, 9].

The barium titanate ($BaTiO_3$) ceramics are classic functional materials for ferroelectric capacitors [10]. Nowadays, the hot-pressure sintering (HPS) techniques, which are well-developed for $BaTiO_3$, require high sintering temperatures (more than 1200 – 1400°C) and long duration (not less than several hours [11]). Despite large energy consumption, the obtained dielectric parameters of the HPS BaTiO$_3$ ceramics are often insufficient for the state-of-the-art applications [12].

To minimize energy consumption and sintering time, one can use the spark-plasma sintering (SPS) method [13]. The SPS enables us to achieve colossal dielectric permittivity, high density and fine crystalline structure of the BaTiO$_3$ nano-grained ceramics [14, 15]. During the SPS the electric current flows between the BaTiO$_3$ nanoparticles in the graphite die. Earlier we have shown that presence of the conducting graphite inclusions, which concentration varies from the surface of the ceramic tablet down to its depth, influences the electrophysical properties of ceramics [16].

In this work we study the electric charge transport and dielectric properties of the BaTiO$_3$ ceramics obtained by SPS method at 1100°C in vacuum under uniaxial pressure of 50 MPa with different carbon content.

The paper has the following structure: section 2 contains information about synthesis of the ceramic samples and their characterization by the scanning electron microscopy (SEM). Section 3 discusses experimental measurements of resistivity and dielectric permittivity temperature dependences, their model explanation, electric field and frequency dependences of resistivity and permittivity and their interrelation measured in the DC and AC regimes for SPS BaTiO$_3$ samples with different carbon content. Section 4 resumes the obtained results.

## 2. SYNTHESIS AND CHARACTERIZATION OF BaTiO$_3$ CERAMICS
### A. Synthesis of BaTiO$_3$ ceramics

All samples were fabricated by the same technique: the mixture of barium titanate and carbon was placed into a graphite press-form with the diameter of 20 mm and sintered during several minutes by the SPS at 1100 °C in vacuum and under uniaxial pressure of 50 MPa.

The **sample #1** was prepared from the 10-gram mixture of 99.5 vol.% of BaTiO$_3$ nanoparticles and 0.5 vol.% of carbon. Sintering time is 6 min. The sample height is 5.7 mm.

The **sample #2** was prepared from the 5-gram mixture of 99.5 vol.% of BaTiO$_3$ nanoparticles and 0.5 vol.% of carbon. Sintering time is 5 min. The sample height is 2.9 mm.

The **sample #3** was prepared from the 5-gram mixture of 99 vol. % of BaTiO$_3$ nanoparticles + 1 vol. % of carbon. Sintering time is 5 min. The sample height is 2.8 mm.

X-ray diffractograms and TEM images of the used BaTiO$_3$ nanoparticles (average size 24 nm) can be found in Ref.[16]. Also, we used larger particles, which TEM images are shown in **Fig. S1** in **Appendix**.

### B. Determination of the carbon content in the sintered samples

To reveal the carbon content in the sintered samples the techniques based on use of the scanning electron microscope (JSM-35) were applied. In the first approach the investigations were performed in the regimes of slow secondary electrons (SE) and more energetic backward scattered electrons (BSE). In the SE regime the signal considerably depends on a sample relief while in the BSE regime it depends on the average atomic number $Z_{av}$ of material. With this method we can compare the carbon content in different samples. Obtained SE and BSE images of the sample surfaces were compared. From comparison, we determined the places, which can be related to the carbon inclusions, and where a small BSE signal is caused by the sample relief. Then one can approximately estimate a percentage of surface occupied by carbon in different samples and compare them.

Shown in **Fig. 1(a)** and **1(b)** are surface images of the sample #1 obtained in the SE and BSE regimes. Comparison of images shows that the dark formation with size of several μm (indicated by arrows in **Fig. 1(a)**) have substantially less $Z_{av}$. One may assume that they contain a large percentage of carbon. As it is seen, the carbon inclusions are very non-uniformly distributed in the sample.

Enlarged image of the fragment "A" in **Fig. 1(a)** is shown in **Figs. 1(c)-(d)**. These images show that a carbon inclusion is also non-uniform and consists of carbon nanoparticles with the minimal size about 100 nm.

Shown in **Figs. 1(e)-1(f)** is the image of another fragment of the sample #1. The size of carbon agglomerates in this fragment is about 10 μm. It is seen that they occupy larger percentage of surface as compared to **Figs. 1(c)-(d)**. Non-uniform carbon distribution is possibly caused by the method of mixing powders in the dry state. One requires to bring enough energy to separate mixtures into separate nanoparticles, to overcome adhesion force between nanoparticles while in the dry state the external energy dissipates on boundaries between grains and their agglomerates.

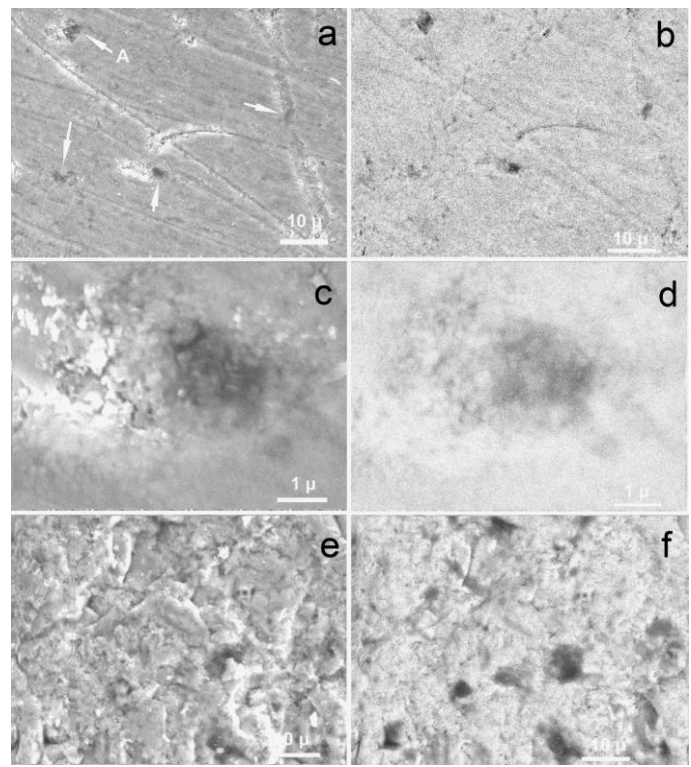

**Figure 1.** SEM images of the sample #1 surface fragment in the SE (**a**) and BSE (**b**) regimes. Enlarged image of the fragment "A" with a carbon inclusion in the sample #1 in the SE (**c**) and BSE (**d**) regimes. SEM images of another surface fragment of the sample #1 in the SE (**e**) and BSE (**f**) regimes.

SEM images of surface fragments of sample #2 are shown in **Fig. 2**. It is seen that carbon inclusions have the size of several μm. In particular, the carbon inclusions/agglomerates shown in **Figs. 2(c)** and **2(d)** are of 10 μm size. These inclusions occupy surface area twice as small as those in the sample #1.

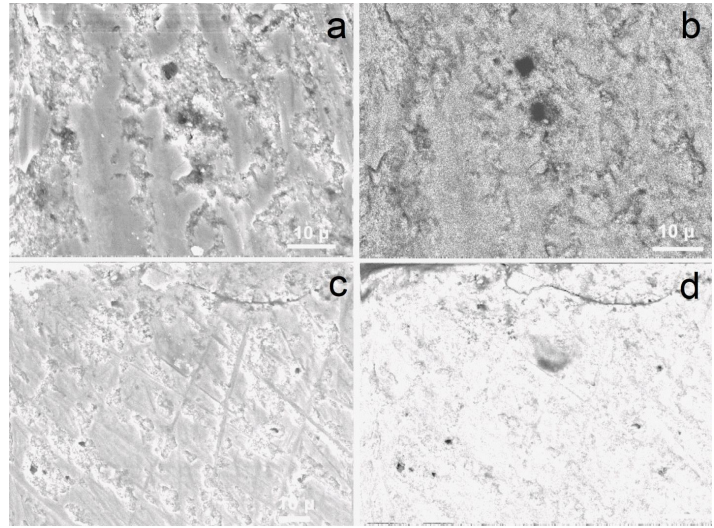

**Figure 2.** SEM images of the sample #2 in the SE **(a, c)** and BSE **(b, d)** regimes.

The SEM images of sample 3 substantially differ from other samples. In contrast to samples #1 and #2 its surface charges strongly under the irradiation by the electron beam and therefore a substantial electric field is induced near the sample surface. Due to the field the focusing of electron beam and brightness of the SEM images appeared unstable, and we cannot obtain a clear image. This peculiarity enables us to conclude that sample #3 has a very small conduction, that could occur due to the small amount of carbon and its very non-uniform distribution in the sample. To clarify the situation, additional studies were carried out. To minimize charging of the sample surface a 30 - 50 nm thick carbon layer was deposited on the surface of the sample #3 by vacuum evaporation, which could not substantially influence the results. Obtained in this way SEM images of the sample #3 surface are shown in **Fig. 3**. It is seen from **Figs. 3(a)-(d)**, that only several inclusions/agglomerates of the 1 - 2 μm size, which have a low contrast, may be related to carbon agglomerates. The low contrast evidences the small carbon content in these agglomerates.

As one can see in **Figs. 3(e, f),** only several inclusions of 200-500 nm size, whose contrast is low, may be related to carbon. The majority of dark spots are related to surface relief. All SEM images of sample #3 confirm that the carbon content in this sample is considerably smaller as compared to samples #1 and #2.

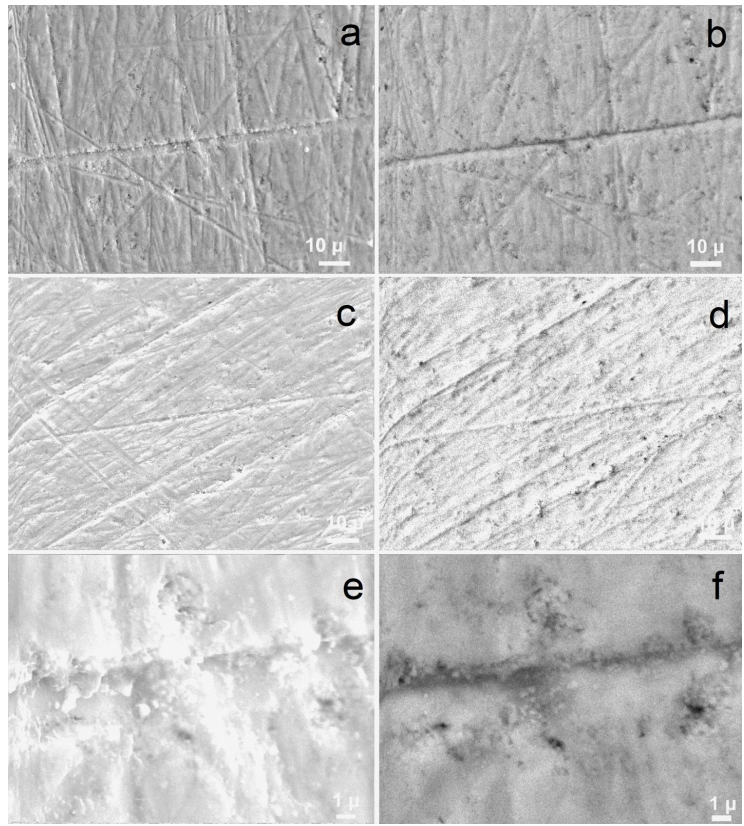

**Figure 3.** SEM images of the sample #3 surface in the SE **(a, c, e)** and BSE **(b, d, f)** regimes.

The BSE images do not allow us to determine the carbon content directly. Accounting for the fact that all samples were mixed and sintered by the same technique, one may suggest that percentage of carbon, which resulted in agglomerates of the μm size, is the same in different samples. Therefore, determination of the agglomerates content in different samples enables us to compare approximately the total content of carbon in the samples. For this purpose, we determined the percentage of surface area occupied by the carbon agglomerates for each SEM image. Based on suggestion that carbon agglomerates have a hemispherical shape (at least in average) we reckon this to obtain a surface concentration of carbon and calculated the bulk concentration. Also, we assumed that the carbon agglomerates are distributed in the sample bulk as shown in the SEM images. This assumption does not have sufficient justification (as well as the previous one). Besides this, determination about which dark spot are related to carbon and which are related to relief is controversial.

Using the above-mentioned assumptions, we estimated the following volume content of carbon in the samples: 0.3 vol.% for sample #1, 0.1 vol.% for sample #2, and 0.005 vol.% for sample #3. Next, assuming that the density of carbon and $BaTiO_3$ being the same as in the bulk material, we estimated the mass content of carbon in the samples: 0.023 wt.% for sample #1, 0.008 wt.% for sample #2, and 0.0004 wt.% for sample #3.

We also used another method enabling us to estimate the carbon content. For this purpose, we recorded the surface image in the regime of Z-contrast at small magnifications. Considering that the distribution of secondary electrons by angles is described by the cosinusoidal function of the angle between the normal to a sample surface and direction of an electron output, the BSE signal of electron detector JSM-35 depends significantly on slope of the sample surface and collects electrons within the angle about 15 degrees from both sides from the normal to the detector. To decrease the contribution of the surface roughness on the BSE signal, the samples were polished by polishing paper with a small grain size. After polishing the samples were cleaned by ultrasonic treatment in distilled water. The amplitude of the BSE signal along the polished part of the sample with the width from 10 to 20 lines was determined from the image, that gave us a more integral signal. As the reference material with a known $Z_{av}$ we used a gold foil of 1 mm width and 30-40 μm thickness. Plasticity of the foil enabled us to press it tightly to the sample surface which decreased the error of measurements. The sample with the foil is shown in **Fig. 4.** The $Z_{av}$ value of the sample was calculated from the ratio of the sample and foil BSE signals. The carbon content in a sample was determined from it.

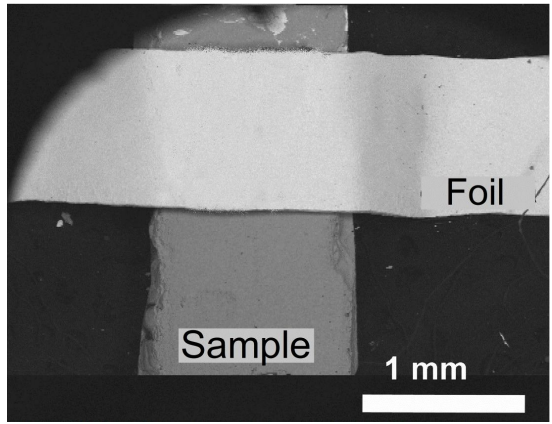

**Figure 4.** SEM images of the sample and a gold foil obtained in the BSE.

Note, that adding 1 atomic % of carbon changes $Z_{av}$ to 33.91, while it is 33.94 for the pure $BaTiO_3$, i.e., $Z_{av}$ value changes to about 0.1 %. Unfortunately, we cannot record the small change using the JSM-35 detector of electrons; only comparative analysis is possible. In this way we estimated that the carbon content is too small in the sample #3, then in atomic percentage it is 0.001 at.%, 2 at.% in the sample #2 and about 6 at.% in the sample #1. In mass percentage this is 0.5 wt.% in the sample #1, 0.15 wt.% in the sample #2 and 0.00008 wt.% in the sample #3.

Considerably smaller values, obtained by the first method, may be explained by that only carbon in the form of μm-sized agglomerates was accounted, while the carbon nanoparticles were distributed inside the samples as well (considered in the second method).

## 3. ELECTRIC TRANSPORT MEASUREMENTS
### A. Experimental description

Studies of electric transport mechanisms were performed in the DC and AC regimes. In the DC regime the principal circuit consisted of a sample serially connected with a load resistor. The voltage was applied by a programmable power supply GW Instek PSP-603. The voltage drop across the sample and load resistor was measured by the digital multimeters Keithley-2000, across the latter to determine electric current through the sample. Measurements of the resistivity were performed in the temperature range from 77 to 290 K in the regime of negligibly small Joule heating. The maximal power dissipated in the samples bulk was in the limits of 0.1 to 10 mW. Also, we performed measurements of the current-voltage characteristics at fixed temperatures in the range from 77 to 290 K in the pulsed regime to minimize the Joule heating. The voltage pulses were applied to the samples by the voltage supply G5-63. The waveforms of voltage drop across the sample and load resistor were measured by the digital oscilloscope Velleman PCS500. To eliminate the impact of transient process at the beginning of the pulse, caused by high resistance of the samples and circuit "parasitic" capacitance of the current through a sample, the parts of waveforms at 9 µs after a pulse switch on were used.

In the AC regime we measured the dependences of resistivity and capacitance at frequencies 10 Hz and 100 kHz in the temperature range 77 to 408 K. The amplitude of the testing AC voltage was small (1 V) to minimize the Joule heating. The AC characteristics were measured by the LCR-meter LCX200 ROHDE & SCHWARZ, when the sample was connected directly without a load resistor. The dielectric permittivity was recalculated from the capacitance measurements.

### B. Electric conduction mechanism in the DC regime

The temperature dependence of samples resistivity demonstrates an activation behavior, which is relevant to the variable range hopping (VRH) conduction obeying the Mott law [17]:

$$\rho \sim exp\left(\frac{T_0}{T}\right)^{1/4}, \qquad T_0 = \frac{21}{N(E)a^3}, \qquad (1)$$

where $N(E)$ is the density of localized electron states in the vicinity of the Fermi level, $a$ is the localization radius of a charge carrier wave function. The DC temperature dependences of resistivity for the samples #1 - #3 are shown in **Fig. 5**. It is seen that the resistivity of samples increases by magnitude under equal temperatures with decreasing the content of carbon, as it was determined by the Z-contrast measurements. The parameter $T_0^{1/4}$ changes from 90 to 135 K$^{1/4}$. This corresponds to the decrease in factor $N(E)a^3$ from $4.8 \cdot 10^{-3}$ to $10^{-3}$ 1/eV, that, as it is shown below, gives the average value $\bar{N}(E)$ of the order of $10^{17}$ 1/(eV·cm³).

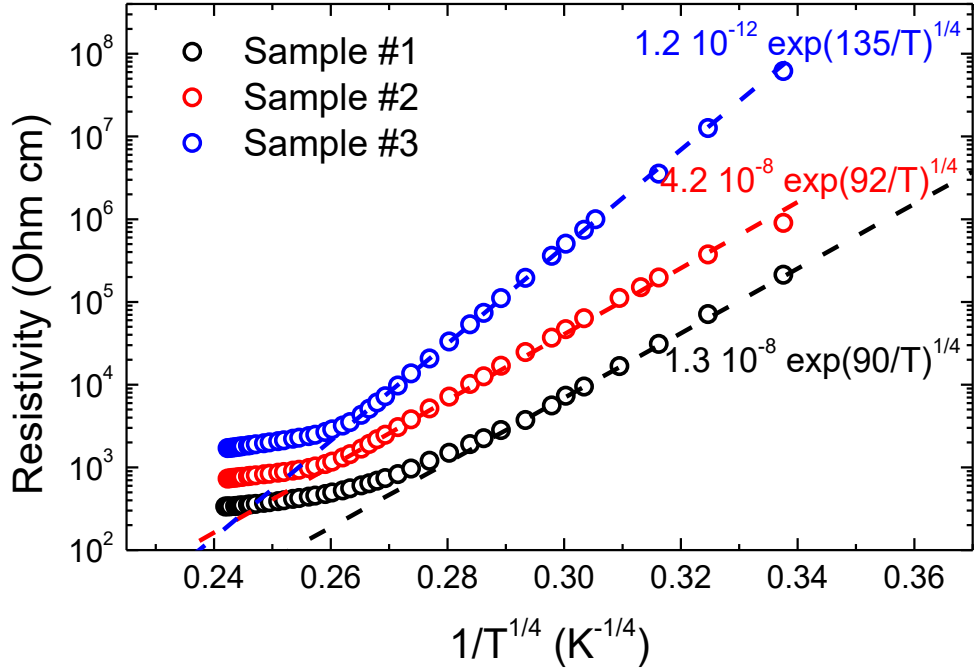

**Figure 5.** Temperature dependences of resistivity of the SPS BaTiO$_3$ ceramics with different carbon content plotted in the Mott law coordinates.

To analyze the observed electric hopping conduction mechanism in more detail, we studied the dependences of resistivity vs electric field strength at different temperatures in the range from 77 to 290 K.

According to Refs. [18, 19], under the condition $eEr_m > k_B T$, the conductivity obeys the law:

$$\sigma(E) \sim exp\left(C\frac{eEr_m}{k_B T}\right), \qquad r_m = a\xi_C/2, \qquad \xi_C = \left(\frac{T_0}{T}\right)^{1/4}, \qquad (2)$$

where $E$ is the electric field strength, $r_m$ is the maximal hopping length along the percolation path, $e$ is the electron charge, $k_B$ is the Boltzmann constant. $C$ is a constant of the order of unity that is equal to 0.18 according to Ref. [18] and 0.8 according to Ref. [19].

The dependences of conductivity vs electric field strength measured for the sample #2 at temperatures of 77 and 150 K are shown in **Fig. 6**. The dependences measured at other temperatures demonstrate similar behavior. It is seen from the dashed lines in **Fig. 6** that experimental dependences are well described by the formula (2). At the same time, the electric field, under which the conductivity becomes temperature-independent and depends only on the electric field strength [20], is not reached in the studied range of electric fields.

The value $T_0^{1.4} = 92$ $K^{1/4}$ for the sample #2. The slope of the curves in **Fig. 6** are $2 \cdot 10^{-4}$ cm/V for $T = 77$ K and $1.1 \cdot 10^{-4}$ for $T = 150$ K. Hence the value $\xi_c$ equals to 31 at 77 K and 26.3 and 150 K, respectively. The maximal hopping length is about 10 nm (by an order of magnitude) at $C = 0.18$ in the weak-field regime. Hence, the estimates of the localization radius are 1.5 and 1 nm at these temperatures. The localization radius is 6.3 nm at 77 K and 4.7 nm at 150 K for C=0.8. Thus, the localization radius of charge carriers wave function equals 1 nm through 6 nm in the samples studied, which is a reasonable value for considered systems. This implies that the density of electron states near the Fermi level agrees by order of magnitude with the above made estimate.

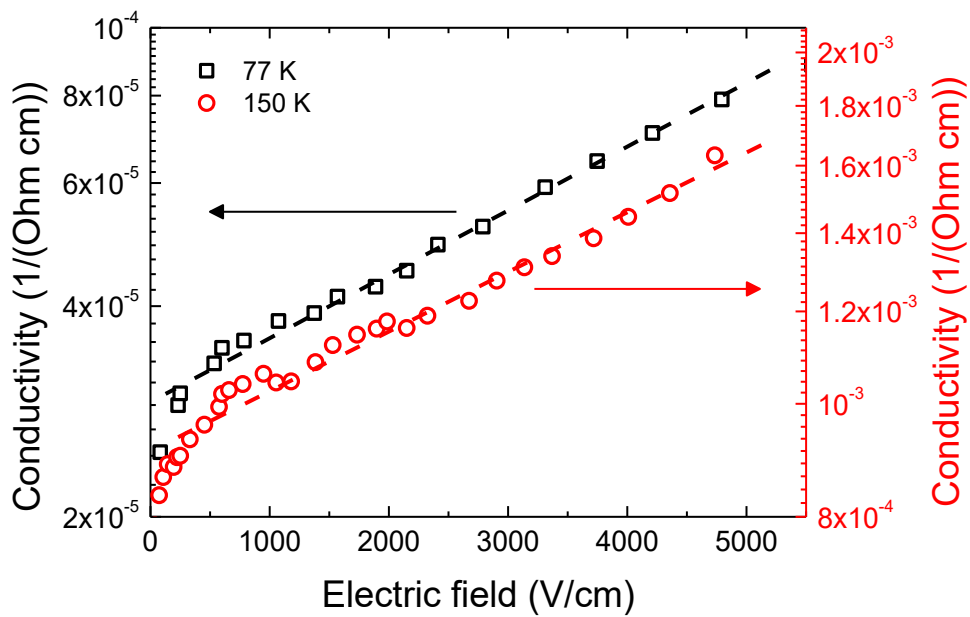

**Figure 6.** Dependence of conductivity vs electric field strength of the sample #2 at temperatures 77 K and 150 K.

To conclude, the low-temperature dependence of resistivity of the SPS BaTiO$_3$ ceramics with carbon, measured in the DC regime, obeys the Mott law at $T < 200$ K and is determined by the VRH conduction mechanism. The obtained temperature and electric field dependences of the VRH conductivity enabled us to determine the localization radius of the charge carriers wave function and density of localized states participating in conduction.

### C. Electric conduction in the AC regime

Shown in **Fig. 7** are the resistivity vs temperature dependences of the samples measured in the AC regime at frequencies $f =$ 10 Hz and 100 kHz. For comparison we show also the temperature dependence of resistivity for the sample #1 measured in the DC regime. It is seen that the dependences remain their activation type conduction in the AC regime. The dependence at 10 Hz, as is seen from **Fig. 7** for the sample #1, is very close to that measured in the DC regime. With increasing frequency, the resistivity decreases strongly (almost by an order of magnitude); it decreases especially strongly at low temperatures. In contrast to the DC regime, the resistivity curves in the AC regime tend to saturate and become almost horizontal at the lowest temperatures. This may be caused by the inhomogeneous distribution of carbon inclusions/agglomerates in the bulk of the samples and by formation of the filaments, which effectively shunt the whole bulk in the AC regime (especially at high frequencies).

At temperatures above the room temperature (300 K) one observes a sharp increase in the slope of resistivity vs temperature dependence. The nature of this peculiarity may be explained by the temperature dependences of the dielectric permittivity shown in **Fig. 8**.

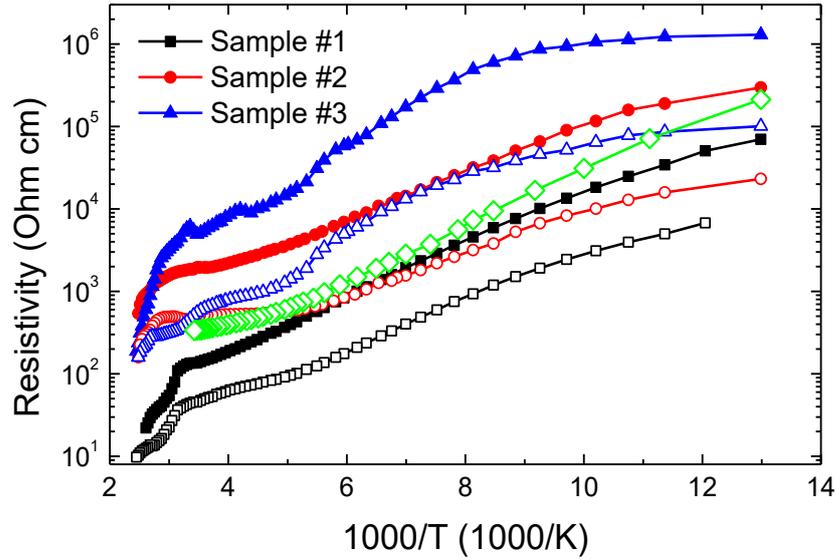

**Figure 7.** Temperature dependences of resistivity measured in the AC regime in the Arrhenius coordinates. Filled symbols correspond to the frequency $f = 10$ Hz, empty symbols correspond to $f = 100$ kHz. Empty light-green diamonds show the temperature dependence of resistivity of the sample #1 in the DC regime.

It is seen from **Fig. 8**, the dielectric permittivity sharply increases and reaches maximum in the temperature region, where one observes a sharp decrease of resistivity. Also, note the correlation between the dielectric permittivity and resistivity for different samples in the low temperature region. The samples with smaller permittivity have higher resistivity. This fact enables us to use the Heywang model [21] for analysis of resistivity. Hence, the difference in resistivity of studied samples is determined by the difference in the carbon impurity concentration and in their dielectric permittivity. In latter case the difference in the dielectric permittivity may be caused, in turn, by difference in the carbon impurity concentration.

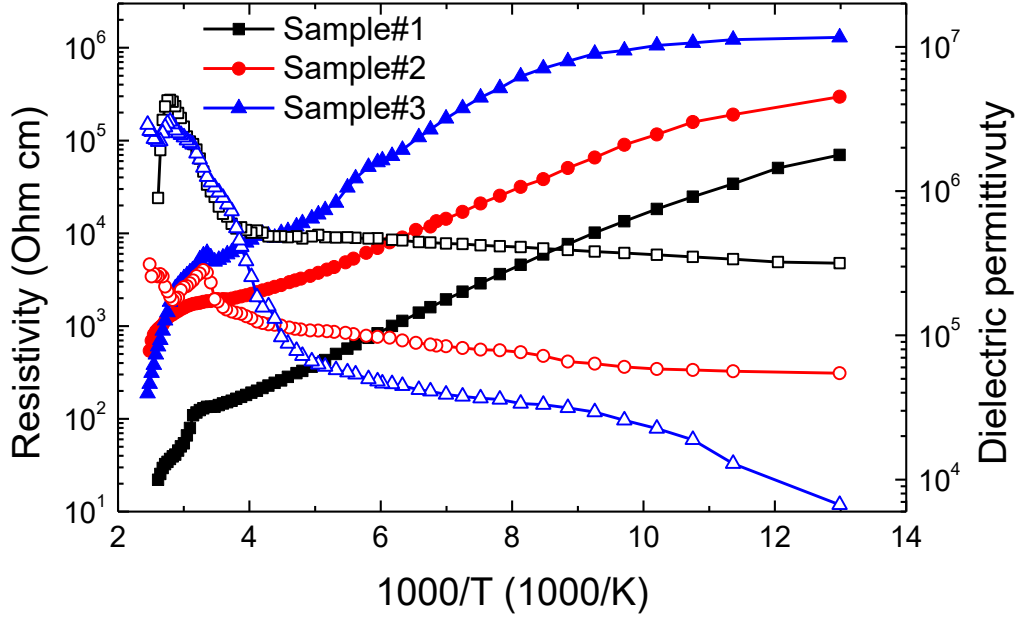

**Figure 8.** Temperature dependences of resistivity (filled symbols) and dielectric permittivity (empty symbols) for the samples #1 - 3.

In general, the observed colossal dielectric constant is characteristic of heterogeneous ferroelectric ceramics. Thus, as it was shown in the works [22, 23], that the colossal dielectric constant of heterogeneous ferroelectric composites and/or ceramics is usually caused by mesoscopic inhomogeneities in electrical conductivity, for example, between grains and grain boundaries, known as the effect of internal barrier layer capacitance (IBLC), as well as by inhomogeneous layers between electrodes and the sample, known as the effect of surface barrier layer capacitance (SBLC). The IBLC effect causes a colossal dielectric constant at low frequencies if the conductivity of the substance between grains is significantly lower than the conductivity of the grains. In this case, the percolation of components with low conductivity (i.e., substances between ceramic grains) occurs and is described by the effective medium approach (EMA), discussed in Refs. [24, 25], as well as in the works of Petzelt et al. [26] and Richetsky et al. [27].

Considering that the EMA is applicable, we may use qualitatively the expression for a localization radius in the effective mass approximation

$$a = \frac{\hbar^2 \varepsilon}{me^2}, \quad (3)$$

where $\hbar$ is the reduced Plank constant, $\varepsilon$ is the effective dielectric permittivity, $m$ is the effective mass of a charge carrier, $e$ is the electron charge. Hence, we obtain

$$T_0 = \frac{21m^3e^6}{N(E)\hbar^6\varepsilon^3}, \qquad \rho \sim exp\left(\frac{A}{T^{1/4}\varepsilon^{3/4}}\right), \qquad A = \frac{m^{3/4}e^{3/2}}{N(E)^{1/4}\hbar^{3/2}}. \qquad (4)$$

Thus, we may expect approximately the inverse dependence between the resistivity $\rho$ and the dielectric permittivity $\varepsilon$ at fixed temperatures in the low temperature range, namely $ln(\rho) \sim 1/\varepsilon^{3/4}$. Of course, some changes may occur also in the density of localized states at the Fermi level. To verify this, additional studies are required.

# 4. CONCLUSION

The conductivity of the SPS BaTiO$_3$ ceramic in the temperature range from 290 K to 77 K is determined by the variable range hopping (VRH) mechanism and obeys the Mott law at small voltages, without charge carriers heating. The current-voltage characteristics measured in the short-pulse regime to avoid the Joule heating, also obey the well-known VRH model for conduction in the strong electric field evolved by Hill [18], Pollak and Riess [19]. Based on these models we estimated the parameters of the VRH conduction: the average value of the localized states density near the Fermi level is of order of 10$^{17}$ eV$^{-1}$cm$^{-3}$, the charge carrier wave function localization radius in average of 1 to 6 nm.

Temperature dependences of resistivity for the samples studied in the AC regime also reveal the behavior inherent to the hopping conduction. The magnitude of resistivity decreases significantly (about by order of magnitude) with increase in the frequency from 10 Hz to 500 kHz.

The important peculiarity is the dependence both of resistivity and dielectric permittivity of SPS BaTiO$_3$ ceramics on the carbon content in samples. In the temperature range of VRH conduction the larger is the carbon content, the smaller is the resistivity and the larger is the dielectric permittivity. Moreover, one may note more or less accurate reciprocal functional relation between the resistivity and the dielectric permittivity.

One must also note a strong increase in the dielectric permittivity at higher temperatures. The dielectric permittivity sharply increases with growing temperature forming a maximum which may be the evidence of a phase transition. At the same time, one does not observe any correlation between the magnitude of the dielectric permittivity maximum and the carbon content, which remains unclear. In the temperature range of this maximum there is observed a sharp increase of the slope of the temperature dependence of resistivity; and the resistivity decreases strongly with growing temperature. Along with reciprocal functional ratio between the resistivity and the dielectric permittivity in the VRH conduction temperature range the sharp increase of resistivity vs temperature slope in the region of dielectric permittivity maximum evidence of the impact of the Heywang model [21] into the conduction mechanism, which manifests itself in changing parameters of conductivity due to the dielectric permittivity changes.

**Authors' contribution.** O.S.P. formulated the idea of the study. O.S.P., D.O.S. T.O.K. and O.V.B. prepared the samples for measurements and performed electrophysical measurements. O.S.P. and V.V.V. analyzed the results of electrophysical measurements and prepared figures. S.E.I., B.P. and V.K. synthesized the ceramic samples. L.D. performed TEM measurements, V.I.S. performed SEM measurements and analyzed the results. V.M.P. coordinated the study and analyzed the results. O.V.B., V.V.V., V.I.S. and A.N.M. wrote the manuscript parts.

**Acknowledgments.** The work of O.S.P., D.O.S. and A.N.M. are funded by the National Research Foundation of Ukraine (project "Manyfold-degenerated metastable states of spontaneous polarization in nanoferroics: theory, experiment and perspectives for digital nanoelectronics", grant application 2023.03/0132). The work of O.V.B. is funded by the NAS of Ukraine, grant No. 07/01-2025(6) "Nano-sized multiferroics with improved magnetocaloric properties." The samples preparation and characterization (S.E.I.) are sponsored by the NATO Science for Peace and Security Programme under grant SPS G5980 "FRAPCOM". L.D. acknowledges support from the Knut and Alice Wallenberg Foundation (grant no. 2018.0237) for TEM research. A part of electrophysical measurements (V.N.P. and V.V.V) are sponsored by the Target Program of the National Academy of Sciences of Ukraine, Project No. 4.8/23-p.

**Appendix A. Materials Characterization and Methods of Sintering**

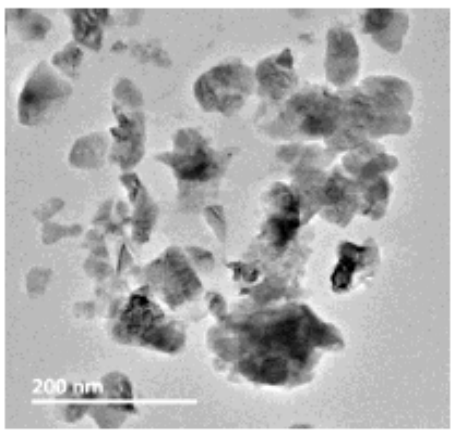

**Figure S1.** TEM images of larger $BaTiO_3$ nanoparticles. TEM images of 24-nm nanoparticles used in the SPS of ceramics are shown in Ref. [16].

Barium titanate ($BaTiO_3$) nanopowder (NanotechCenter LLC, Ukraine) with a mean particle diameter of 24 nm, specific surface area of 25 m²/g served as the ceramic matrix phase. Nanosized spherical carbon powder was used as the conductive additive. $BaTiO_3$ and carbon powders were dry-mixed in sealed 50 ml lab tube using an Intelli-Mixer RM-1M (ELMI, Latvia) for 24 hours. The mixing protocol consisted of continuous 110° bidirectional rotations with small-amplitude oscillations (1.5 s per cycle).

Samples were consolidated using a Spark Plasma Sintering Furnace HP D 25 (FCT Systeme GmbH, Germany). Sintering was carried out in a graphite mold with a diameter of 20 mm, using graphite foil to protect the mold. In the same press form in which sintering took place, $BaTiO_3$-C mixture was pre-pressed under a load of 5 kN (16 MPa). Before starting to sinter, a vacuum atmosphere was created in the chamber, and a load of 16 kN (50 MPa) was applied. Next, heating was carried out to a temperature of 1100C at a heating rate of 400C/min, with exposure at this temperature for 5 min. After the exposure time is over, the heating is turned off, and the load is gradually removed to 5 kN (16 MPa) within 3 minutes, the temperature of the sample currently is about 600C. Next, the chamber is filled with nitrogen (N2) and the subcooling is finally removed, free cooling to room temperature occurs. The sintering mode of SPS of the BaTiO3 sample is shown in **Figure S1**.

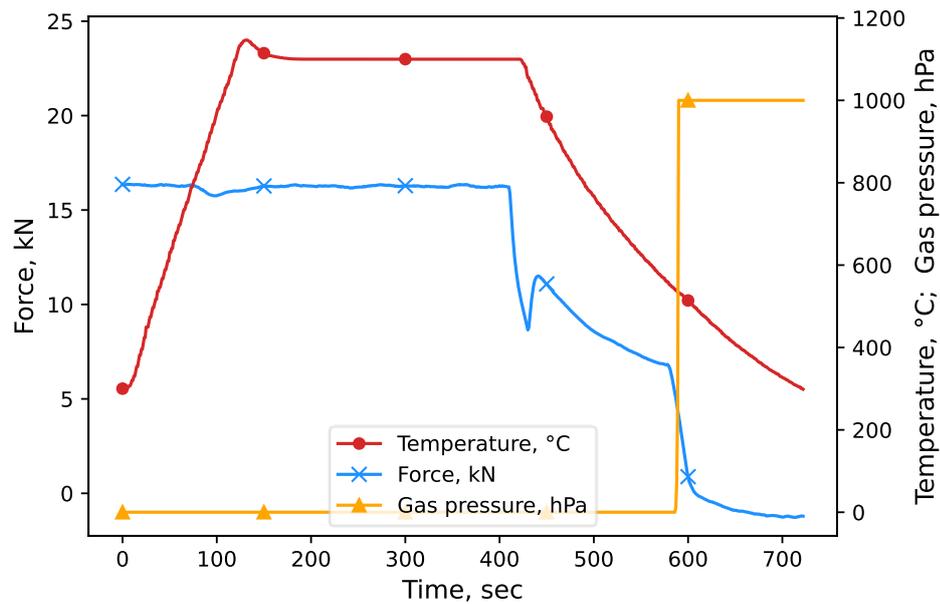

**Figure S1.** Sintering mode of the SPS BaTiO$_3$ ceramics.

Geometric densities were calculated from sample mass and dimensions (see **Table S1**). BaTiO$_3$ density for calculation was taken 6.02 for BaTiO$_3$ and 2 for C. Relative sample densities ranged from 91.30% to 94.56%, indicating 5-8% porosity typical for SPS-processed ceramics [16]. The small density differences between samples 1 and 2 (0.1 g/cm³) can be explained by slightly longer holding time of the sample 1 (6 min vs 5 min). Notably, sample 3 (prepared with 1.0 vol% nominal carbon) exhibited the highest density (5.68 g/cm³), contradicting the expected trend of decreasing density with increasing carbon content.

**Table S1.** Samples properties

| Sample | Nominal carbon, vol% | Nominal carbon, wt% | Mixture mass, g | Holding time, min | Height, mm | Volume, cm³ | Geometric density, g/cm³ | Relative density, % |
|---|---|---|---|---|---|---|---|---|
| 1 | 0.5 | 0.167 | 10 | 6 | 5.7 | 1.79 | 5.59 | 92.96 |
| 2 | 0.5 | 0.167 | 5 | 5 | 2.9 | 0.91 | 5.49 | 91.30 |
| 3 | 1.0 | 0.334 | 5 | 5 | 2.8 | 0.88 | 5.68 | 94.56 |